\documentclass[epj]{webofc}
\usepackage[varg]{txfonts}   
\usepackage{graphicx}
\usepackage{color} 
\usepackage{epsfig}
\usepackage{bm}       
\usepackage{latexsym}
\usepackage{amsmath}
\usepackage{amsfonts}
\usepackage{amssymb}
\woctitle{21st International Conference on Few-Body Problems in Physics}
\begin{document}
\title{Strong diquark correlations inside the proton}
\author{
     Jorge Segovia\inst{1}\fnsep\thanks{\email{segonza@usal.es}} 
}

\institute{
Instituto Universitario de F\'isica Fundamental y Matem\'aticas
(IUFFyM), \\ Universidad de Salamanca, E-37008 Salamanca, Spain 
}

\abstract{
Quantum Chromodynamics is thought to be the relativistic quantum field theory 
that describes the strong interaction of the Standard Model. This interaction 
produces mesons but it is also able to generate quark-quark (diquark) 
correlations inside baryons. In this work, we employ a continuum approach to 
QCD based on Dyson-Schwinger equations to calculate the electromagnetic form 
factors of the proton and analyze in a deeper way the consequences of having 
strong diquark correlations. Comparison with the experimental data reveals 
that the presence of strong diquark correlations within the proton is 
sufficient to understand empirical extractions of the flavour-separated form 
factors. The explained reduction of the ratios $F_{1}^{d}/F_{1}^{u}$ and 
$F_{2}^{d}/F_{2}^{u}$ at high $Q^{2}$ in the quark-diquark picture are 
responsible of the precocious scaling of the $F_{2}^{p}/F_{1}^{p}$ observed 
experimentally.
}
\maketitle
%

\vspace*{-0.20cm}
\section{Origin of diquarks}
\label{sec:diquarks}

Quantum Chromodynamics (QCD) is the strong-interaction part of the Standard 
Model of Particle Physics and solving QCD presents a fundamental problem that 
is unique in the history of science. Never before we have been confronted by a 
theory whose elementary excitations are not those degrees of freedom readily 
accessible via experiment, that is, whose elementary excitations are confined. 
Moreover, QCD generates forces which are so strong that they produce about 
$98\%$ of the mass of the proton and, consequently, most of the mass in the 
visible Universe. The underlying phenomenon which generates this 
mass -- from nothing -- is dynamical chiral symmetry breaking (DCSB). Neither 
confinement nor DCSB are apparent in QCD's Lagrangian and yet they play the 
dominant role in determining the observable characteristics of real-world QCD, 
{\it e.g.}, hadron masses and coupling.

An important consequence of DCSB is that any interaction capable of creating 
bound-states of a light dressed-quark and -antiquark will necessarily also 
generate strong colour-antitriplet correlations between any two dressed quarks 
contained within a nucleon. Although a rigorous proof within QCD cannot be 
claimed, this assertion is based upon an accumulated body of evidence, gathered 
in two decades of studying two- and three-body bound-state problems in hadron 
physics (see, for instance, 
Refs.~\cite{Cahill:1987qr,Bender:1996bb,Cloet:2008re,Segovia:2014aza, 
Segovia:2015hra}). No realistic counter examples are known; and the existence 
of such diquark correlations is also supported by simulations of 
lattice QCD~\cite{Alexandrou:2006cq,Babich:2007ah}.

The properties of diquark correlations have been charted. Diquarks are 
confined. Additionally, owing to properties of charge-conjugation, a diquark 
with spin-parity $J^P$ may be viewed as a partner to the analogous $J^{-P}$ 
meson~\cite{Cahill:1987qr}. It follows that scalar, isospin-zero and 
pseudovector, isospin-one diquark correlations are the strongest; and whilst no 
pole-mass exists, the following mass-scales, which express the strength and 
range of the correlation and are each bounded below by the partnered meson's 
mass, may be associated with these 
diquarks~\cite{Cahill:1987qr,Alexandrou:2006cq,Babich:2007ah}: 
$m_{[ud]_{0^+}} \approx 0.7-0.8\,{\rm GeV} $, $m_{\{uu\}_{1^+}}  \approx 
0.9-1.1\,{\rm GeV}$, with $m_{\{dd\}_{1^+}}=m_{\{ud\}_{1^+}} = 
m_{\{uu\}_{1^+}}$ in the isospin symmetric limit. Realistic diquark 
correlations are also soft. Finally, they possess an electromagnetic size that 
is bounded below by that of the analogous mesonic system, \emph{viz}. 
\cite{Maris:2004bp,Roberts:2011wyS}: $r_{[ud]_{0^+}} \gtrsim r_\pi$, 
$r_{\{uu\}_{1^+}} \gtrsim r_\rho$, with $r_{\{uu\}_{1^+}} > r_{[ud]_{0^+}}$.


\begin{figure}[!t]
\hspace*{0.50cm}
\includegraphics[clip,width=0.4\textwidth,height=0.18\textheight]
{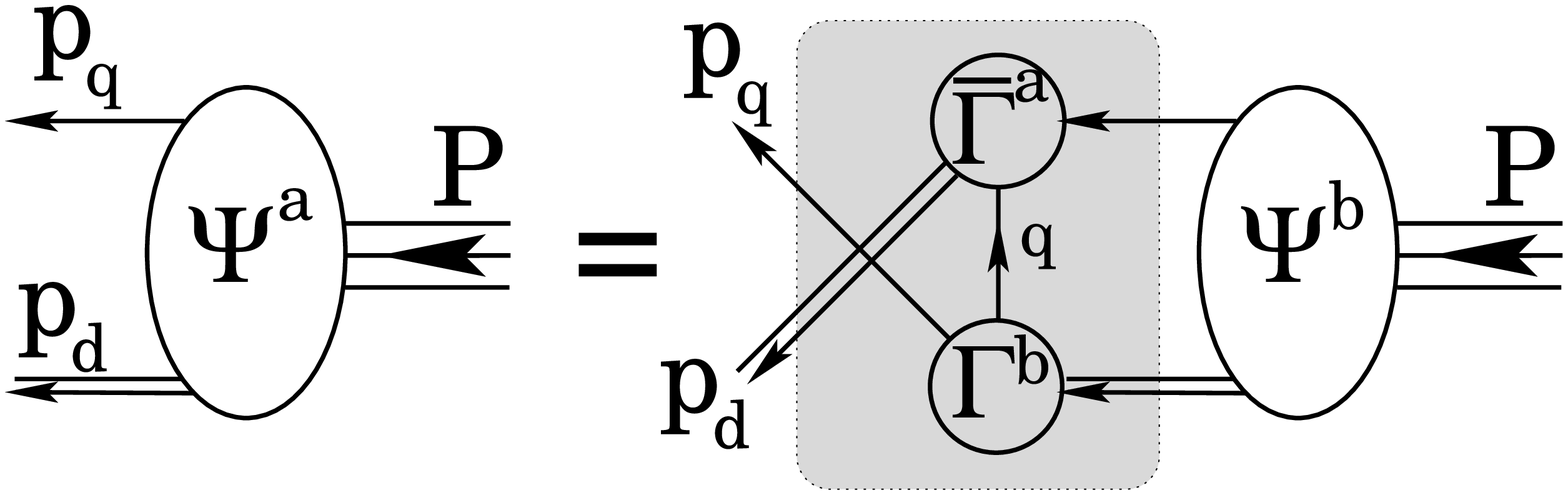}
\hspace*{1.00cm}
\includegraphics[clip,width=0.4\textwidth,height=0.20\textheight]
{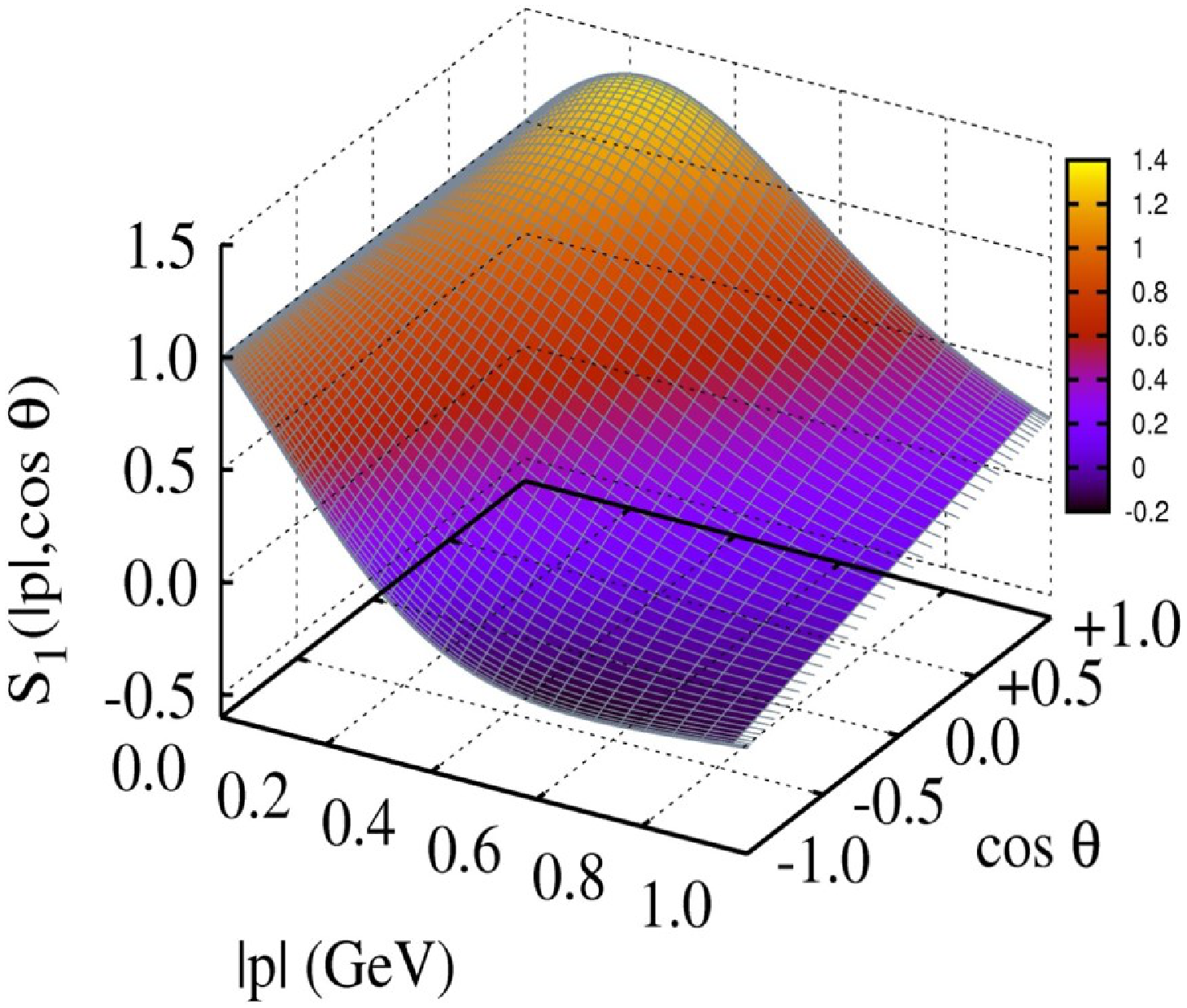}
\caption{\label{fig:Faddeev} {\bf Left panel:} Poincar\'e covariant 
Faddeev equation. $\Psi$ is the Faddeev amplitude for a baryon of total 
momentum $P= p_q + p_d$, where $p_{q,d}$ are, respectively, the momenta of the 
quark and diquark within the bound-state. The shaded area demarcates the 
Faddeev equation kernel: \emph{single line}, dressed-quark propagator; 
$\Gamma$, diquark correlation amplitude; and \emph{double line}, diquark 
propagator.
{\bf Right panel:} Dominant piece in the nucleon's eight-component 
Poincar\'e-covariant Faddeev amplitude: $S_1(|p|,\cos\theta)$. In the nucleon 
rest frame, this term describes that piece of the quark-diquark relative 
momentum correlation which possesses zero \emph{intrinsic} quark-diquark 
orbital angular momentum, \emph{i.e}.\ $L=0$ before the propagator lines are 
reattached to form the Faddeev wave function.  Referring to 
Fig.~\ref{fig:Faddeev}, $p= P/3-p_q$ and $\cos\theta = p\cdot P/\sqrt{p^2 
P^2}$. The amplitude is normalised such that its $U_0$ Chebyshev moment is 
unity at $|p|=0$.
}
\vspace*{-0.60cm}
\end{figure}

\vspace*{-0.10cm}
\section{Diquarks inside the nucleon}
\label{sec:diquarksnucleon}

The existence of tight diquark correlations considerably simplifies analyses of 
the three valence-quark scattering problem and hence baryon bound states. The 
nucleon is a compound system whose properties and interactions are primarily 
determined by the quark$+$diquark structure evident in the Poincar\'e covariant 
Faddeev equation depicted in the left panel of Fig.~\ref{fig:Faddeev}. While an 
explicit three-body term might affect fine details of baryon structure, the 
dominant effect of non-Abelian multi-gluon vertices is expressed in the 
formation of diquark correlations. 

A nucleon (and kindred baryons) described by the left panel of 
Fig.~\ref{fig:Faddeev} can be view as a Borromean bound-state, the binding 
within which has two contributions. One part is expressed in the formation of 
tight diquark correlations. That is augmented, however, by attraction generated 
by the quark exchange depicted in the shaded area of the left panel of 
Fig.~\ref{fig:Faddeev}. This exchange ensures that diquark correlations within 
the nucleon are fully dynamical: no quark holds a special place because each one 
participates in all diquarks to the fullest extent allowed by its quantum 
numbers. The continual rearrangement of the quarks guarantees, \emph{inter} 
\emph{alia}, that the nucleon's dressed-quark wave function complies with Pauli 
statistics.

It is important to highlight that both scalar-isoscalar and 
pseudovector-isotriplet diquark correlations feature within a nucleon. The 
relative probability of scalar versus pseudovector diquarks in a nucleon is a 
dynamical statement. Realistic computations predict a scalar diquark strength 
of approximately $60\%$~\cite{Cloet:2008re,Segovia:2014aza,Segovia:2015hra}. As 
will become clear, this prediction can be tested by contemporary experiments.

The quark$+$diquark structure of the nucleon is elucidated in the right panel 
of Fig.~\ref{fig:Faddeev}, which depicts the leading component of its Faddeev 
amplitude: with the notation of Ref.~\cite{Segovia:2014aza}, 
$S_1(|p|,\cos\theta)$, computed using the Faddeev kernel described therein. 
This function describes a piece of the quark$+$scalar-diquark relative momentum 
correlation. Notably, in this solution of a realistic Faddeev equation there is 
strong variation with respect to both arguments. Support is concentrated in 
the forward direction, $\cos\theta >0$, so that alignment of $p$ and $P$ is 
favoured; and the amplitude peaks at $(|p|\simeq M_N/6,\cos\theta=1)$, whereat 
$p_q \approx P/2 \approx p_d$ and hence the \emph{natural} relative momentum is 
zero. In the antiparallel direction, $\cos\theta<0$, support is concentrated at 
$|p|=0$, \emph{i.e}.\ $p_q \approx P/3$, $p_d \approx 2P/3$.


\begin{figure}[!t]
\centerline{
\includegraphics[clip,width=0.40\textwidth]{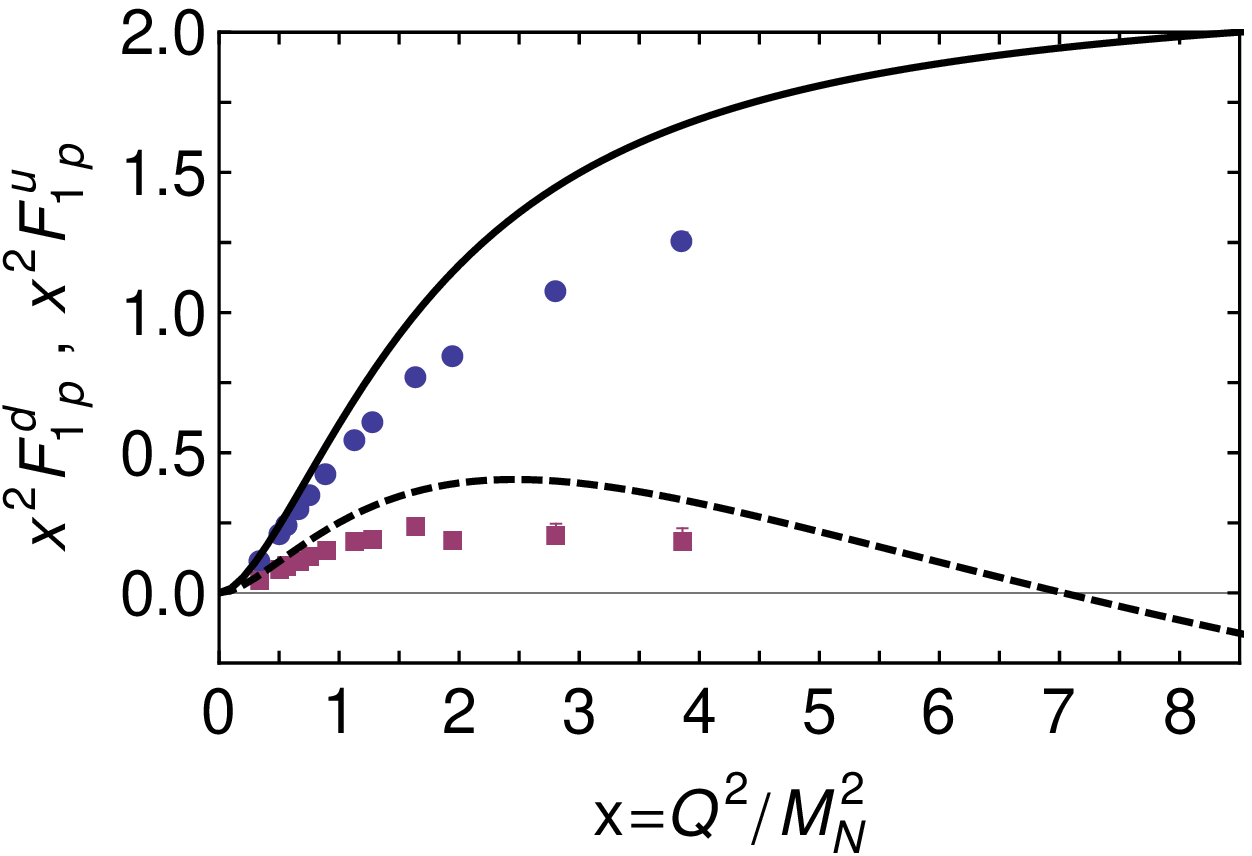}
\hspace*{1.00cm}
\includegraphics[clip,width=0.40\textwidth]{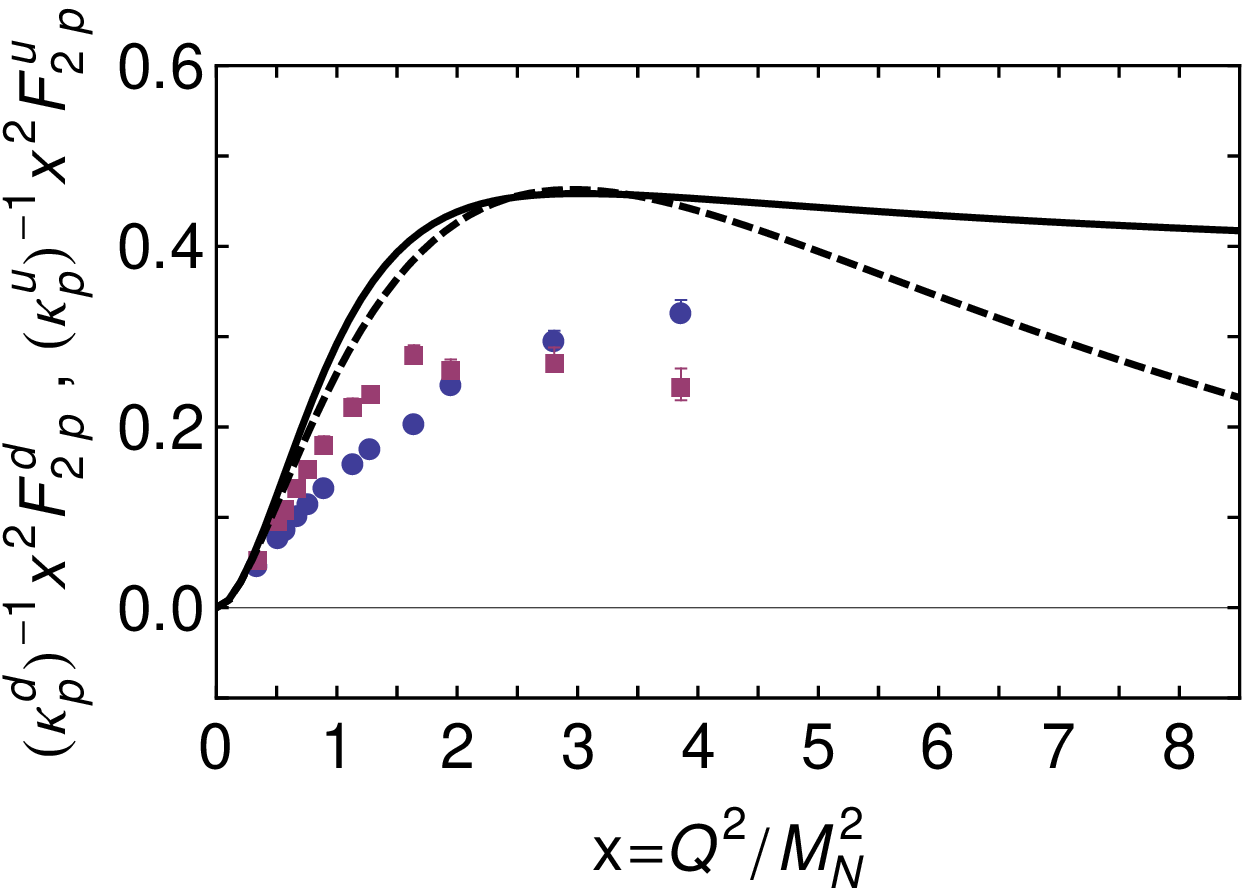}}
\caption{\label{fig:F1F2fla1} {\bf Left panel:} Flavour separation of the 
proton's Dirac form factor as a function of $x=Q^2/M_N^2$.  The 
solid-curve is the $u$-quark contribution, and the dashed-curve is the 
$d$-quark contribution. Experimental data taken from 
Ref.~\protect\cite{Cates:2011pz} and references therein: circles -- $u$-quark; 
and squares -- $d$-quark. {\bf Right panel:} Same for Pauli form factor.
%
}
\vspace*{-0.60cm}
\end{figure}

\vspace*{-0.10cm}
\section{Predictions of diquark clustering}
\label{sec:nucleoncurrent}

A nucleon described by the Faddeev equation in Fig.~\ref{fig:Faddeev} is 
constituted from dressed-quarks, any two of which are always correlated as 
either a scalar or pseudovector diquark. If this is a veracious description of 
Nature, then the presence of these correlations must be evident in many 
physical observables. We focus our attention on the flavour separated versions 
of the Dirac a Pauli form factors of the nucleon. The evaluation within our 
framework of the nucleon electromagnetic current is detailed in 
Ref.~\cite{Segovia:2014aza} and the results we describe herein are derived from 
that analysis.

Figure~\ref{fig:F1F2fla1} displays the proton's flavour separated Dirac and 
Pauli form factors.  The salient features of the data are: the $d$-quark 
contribution to $F_1^p$ is far smaller than the $u$-quark contribution; 
$F_2^d/\kappa_d>F_2^u/\kappa_u$ on $x<2$ but this ordering is reversed on $x>2$; 
and in both cases the $d$-quark contribution falls dramatically on $x>3$ whereas 
the $u$-quark contribution remains roughly constant. Our calculations are in 
semi-quantitative agreement with the empirical data.

It is natural to seek an explanation for the pattern of behaviour in 
Fig.~\ref{fig:F1F2fla1}. We have emphasised that the proton contains scalar 
and pseudovector diquark correlations. The dominant piece of its Faddeev wave 
function is $u[ud]$; namely, a $u$-quark in tandem with a $[ud]$ scalar 
correlation, which produces $62\%$ of the proton's normalisation. If this were 
the sole component, then photon--$d$-quark interactions within the proton would 
receive a $1/x$ suppression on $x>1$, because the $d$-quark is sequestered in a 
soft correlation, whereas a spectator $u$-quark is always available to 
participate in a hard interaction. At large $x=Q^2/M_N^2$, therefore, scalar 
diquark dominance leads one to expect $F^d \sim F^u/x$.  Available data are 
consistent with this prediction but measurements at $x>4$ are necessary for 
confirmation.

\begin{figure}[t]
\centerline{%
\includegraphics[clip,width=0.40\textwidth]{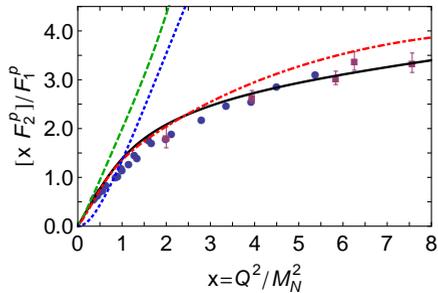}}
\caption{\label{figF1F2fla3} Proton ratio $R_{21}(x) = x F_2(x)/F_1(x)$, 
$x=Q^2/M_N^2$.  Curves:
solid (black) -- full result, determined from the complete proton Faddeev wave 
function and current; dot-dashed (red) -- momentum-dependence of the 
scalar-diquark contribution; dashed (green) -- momentum-dependence of that 
component of the scalar diquark contribution to the proton's Faddeev wave 
function which is purely $S$-wave in the rest-frame; dotted (blue) -- 
momentum-dependence of the pseudovector diquark contribution. The data are 
drawn from Ref.~\protect\cite{Cates:2011pz} and references therein.
%
}
\vspace*{-0.60cm}
\end{figure}

It is natural now to consider the proton ratio: $R_{21}(x) = x F_2(x)/F_1(x)$, 
$x=Q^2/M_N^2$, drawn in Fig.~\ref{figF1F2fla3}. The momentum dependence of 
$R_{21}(x)$ is principally determined by the scalar diquark component of the 
proton. Moreover, the rest-frame $L=1$ terms are seen to be critical in 
explaining the data: the behaviour of the dashed (green) curve highlights the 
impact of omitting these components.


\section{Summary}
\label{sec:summary}

We have explained that the same interaction which produces mesons is able to 
generate tight, dynamical colour-antitriplet quark-quark correlations inside 
baryons. The quark-diquark picture of a baryon should have numerous empirical 
consequences. Focusing on the flavour-separated proton's Dirac and Pauli form 
factors, we have seen that dominant scalar diquark correlations inside the 
proton is enough to observe that, at large $Q^{2}$, the $d$-quark contributions 
to both Dirac and Pauli proton form factors are reduced relative to the 
$u$-quark contributions. Moreover, the explained reduction of the ratios 
$F_{1}^{d}/F_{1}^{u}$ and $F_{2}^{d}/F_{2}^{u}$ at high $Q^{2}$ in the 
quark-diquark picture are responsible of the precocious scaling of the 
$F_{2}^{p}/F_{1}^{p}$ observed experimentally.

\vspace*{-0.10cm}

\begin{acknowledgement}
A significant body of this work is drawn from Ref.~\cite{Segovia:2015ufa}. I am 
grateful for insightful comments from
V.~Mokeev,
I.\,C.~Clo\"et,
R.~Gothe,
and T.-S.\,H.~Lee.
%
I acknowledges financial support from a postdoctoral IUFFyM contract at 
\emph{Universidad de Salamanca}, Spain.
\end{acknowledgement}

\vspace*{-0.10cm}

%
\bibliography{SegoviaJ_21stFew-Body}

\end{document}